\documentstyle[aps,preprint]{revtex}
\begin{document}
\title{Antiferromagnetic ordering in the Kondo lattice system  Yb$_2$Fe$_3$Si$_5$}
\author{Yogesh Singh, S. Ramakrishnan}
\address{Tata Institute of Fundamental Research, Mumbai-400005, India}
\author{Z. Hossain, and C. Geibel}
\address{Max-Planck-Institute for Chemical Physics of Solids, D-01187 Dresden, Germany}
\maketitle
\begin{abstract}
\noindent
Compounds belonging  to the R$_2$Fe$_3$Si$_5$ series exhibit
unusual superconducting and magnetic properties. Although
a number of studies have been made on the first reentrant antiferromagnet
superconductor Tm$_2$Fe$_3$Si$_5$, the physical properties of Yb$_2$Fe$_3$Si$_5$  are largely
unexplored. In this work, we attempt to provide a
comprehensive study of bulk properties such as, resistivity,
susceptibility and heat-capacity of a well characterized polycrystalline
Yb$_2$Fe$_3$Si$_5$. Our measurements indicate that Yb$^{3+}$ moments 
order antiferromagnetically below 1.7~K. Moreover, the system behaves as
a Kondo lattice with large Sommerfeld coefficient ($\gamma$)
of 0.5~J/Yb mol K$^{2}$ at 0.3~K, which is well below T$_N$. The
absence of superconductivity in Yb$_2$Fe$_3$Si$_5$ down to 0.3~K 
at ambient pressure is attributed to the presence of the Kondo effect.
\vskip 1truecm 
\noindent 
Ms number ~~~~~~~~~~~~PACS number:~72.10.Fk, 72.15.Qm, 75.20.Hr, 75.30.Mb\\
\end{abstract}
\newpage
\section{Introduction}
\label{sec:INTRO}
\noindent
Ternary rare earth transition metal silicides, which form in a variety 
of crystal structures have been widely investigated  due to their remarkable 
physical properties \cite{r1,r2}. A few of them also undergo superconducting
transition at low temperatures \cite{r3,r4}. In the early 80's, several investigations
have been carried out to understand the superconductivity and magnetism exhibited by compounds
belonging to the R$_2$Fe$_3$Si$_5$ system \cite{r5,r6,r7,r8}. In this family
the Fe atoms do not carry any magnetic moment but help in building a large density of
states at the Fermi level \cite{r9}. It is now well established that a
member of this series, namely, Tm$_2$Fe$_3$Si$_5$ \cite{r10} is the first
reentrant antiferromagnetic superconductor (perhaps the only one down to 50 mK). 
It is worthwhile to point out that we still do not know why the antiferromagnetic 
order among Tm$^{3+}$ ions is a deterrent to the superconductivity in 
Tm$_2$Fe$_3$Si$_5$ given that the neutron scattering studies \cite{r10} could 
not find any ferromagnetic component in the magnetic unit cell in this compound.
A recent report has suggested  that an antiferromagnet Er$_2$Fe$_3$Si$_5$ 
\cite{r11} (below 2.5~K) also shows superconductivity below 1~K whereas an earlier 
heat-capacity study \cite{r8} indicated  quadruple magnetic transitions without 
any superconductivity down to 1.5~K. A previous study \cite{r8} on an arc melted Yb$_2$Fe$_3$Si$_5$ sample reported the observation of antiferromagnetic
ordering of Yb moments  around 1.7~K.
However, that study did neither establish the Kondo lattice 
behaviour nor the high value of the Sommerfeld coefficient ($\gamma$) as we will 
see later.  In this context, the physical properties of 
Yb$_2$Fe$_3$Si$_5$  are largely unexplored as compared to other compounds 
of the R$_2$Fe$_3$Si$_5$ series. Moreover, compounds with magnetic trivalent
or nearly trivalent Yb are interesting since valence instability and Kondo
interaction can lead to unusual properties. Thus, we need to understand the absence of
superconductivity in Yb compound given that both Tm and Lu of the same series
exhibit superconductivity. In this work, we  provide a comprehensive study of bulk 
properties such as, resistivity, susceptibility, and heat capacity of a well characterized 
polycrystalline sample of Yb$_2$Fe$_3$Si$_5$.
\section{EXPERIMENTAL DETAILS}
\label{sec:EXPT}
\noindent
Samples of Yb$_2$Fe$_3$Si$_5$ were made by melting the individual constituents 
(with 2\% Yb excess and other elements taken in 
stoichiometric proportions) in an Alumina crucible which is placed inside 
a Tantalum container and sealed in an arc furnace under high purity argon
atmosphere. The whole assembly is heated in a vertical furnace. The purity of the
rare-earth element Yb and the transition element Fe were 99.99\% whereas the 
purity of Si was 99.999\%. The X-ray powder diffraction pattern of the sample 
did not show the presence of any parasitic impurity phases. The quality of the
sample was checked with EDAX analysis confirming that the sample is essentially
single phase with 2-3-5 composition. The nearest rare-earth 
distances in Yb$_2$Fe$_3$Si$_5$ is 3.8 $\AA$. 

The temperature dependence of the magnetic susceptibility ($\chi$) was measured using
the commercial Squid magnetometer (MPMS5, Quantum Design, USA) in a field of 1~kOe in the 
temperature range from 1.9 to 300~K. The ac susceptibility was measured using a home 
built susceptometer \cite{r13} from 1.5 to 20~K. The resistivity was measured using a 
four-probe dc technique with contacts  made using silver paint on a bar shaped sample of 
1~mm thick, 10~mm length and 2 mm width. The temperature was measured using a calibrated Si
diode (Lake~Shore~Inc., USA) sensor. The sample voltage was measured with a
nanovoltmeter (model 182, Keithley, USA)  with  a current of 10~mA using a
20~ppm stable (Hewlett Packard, USA) current  source. All the data were
collected using an IBM compatible PC/AT via IEEE-488 interface. The
heat-capacity in zero field between 0.3 to 30~K was measured using an
automated adiabatic heat pulse method. A calibrated germanium resistance
thermometer (Lake~Shore~Inc, USA) was used as the temperature sensor in this
range.
\section{RESULTS}
\subsection{Magnetic susceptibility studies}
\label{sec:SUS1}
\noindent
The temperature dependence of the inverse dc magnetic susceptibility
(1/$\chi_{dc}$) of a polycrystalline Yb$_2$Fe$_3$Si$_5$ in a
field of 1 kOe from 1.9 to 300~K is shown in the Fig.~1. 
The high temperature susceptibility (100~K$<$T$<$300~K) is fitted to a
modified Curie-Weiss expression which is given by,
$$ \chi~=~\chi_0~+~{C \over (T~-~\theta_p)}~, \eqno(1) $$
Here, $\chi_0$ is the temperature independent susceptibility
(sum of Pauli, Landau and core susceptibilities), C is the Curie constant
and $\theta_p$ is the Curie-Weiss temperature.
The Curie constant C can be written in terms of the effective
moment as,
$$ C~=~{\mu_{eff}^2~x \over 8}~, \eqno(2) $$
where $x$ is the concentration of Yb$^{3+}$ ions (x~=~2 for
Yb$_2$Fe$_3$Si$_5$).
 The value of $\chi$(0) is extremely small
(7$\times$10$^{-4}$ emu/mol) while the values of the effective magnetic moment
($\mu_{eff}$) and $\theta_p$ are found to be 4.5 $\mu_B$ and $-$14.4~K, respectively. 
The estimated effective moment is found to be nearly equal to the free ion moment 
of the magnetic Yb$^{3+}$ ion which rules out the possibility of mixed
valent nature of Yb as conjectured in an earlier study \cite{r8}.
One can also observe that below 100~K, the $\chi$ data show
deviation from Curie-Weiss plot which could be due to the combined effect
of crystal field and Kondo lattice contributions. 
The inset shows the ac-susceptibility behavior of the sample at low temperature. This inset clearly
indicates antiferromagnetic ordering of Yb$^{3+}$ moments at 1.7~K.
It must be emphasized here that we do not see the influence of
any impurity contribution to $\chi$(T) and especially the absence
of Yb$_2$O$_3$ (T$_N$=2.3~K) which is generally present in many Yb based compounds.
The presence of a single distinct anomaly in $\chi$ at 1.7~K clearly
suggests the high quality of the sample.

Earlier study \cite{r8} showed that the 
antiferromagetic ordering temperature of the heavy rare-earth R$_2$Fe$_3$Si$_5$ 
compounds do follow the deGennes scaling \cite{r14} [$\sim$(g$_J$-1)$^2$~J(J+1)] 
indicating that the dominant magnetic interaction is the RKKY interaction.
Surprisingly, the same study . But the same study revealed that the antiferromagnetic ordering 
temperature of Yb$_2$Fe$_3$Si$_5$ lies far above the theoretical curve \cite{r8}. 
In our opinion, there is no reason to invoke something else 
other than the RKKY interaction for explaining the relatively large
value of T$_N$. An ordering temperature of 1.7 K is a very
common value for Yb-compounds. A slightly enhanced 4f-hybridization, 
besides inducing some Kondo interaction, leads at first to an increase
of the magnetic ordering temperature.
\subsection{Resistivity studies}
\label{sec:RES1}
\noindent
The temperature dependence of the resistivity ($\rho$) data of Yb$_2$Fe$_3$Si$_5$
is shown in Fig.~2. The two insets (a) and (b) display the $\rho$ behaviour
from 1.4 to 5~K and 1.4~ to 50~K, respectively. The large value of the $\rho$(T) at room temperature 
and a low value of the residual resistivity ratio down  to 5~K could be due to either 
the structural disorder (inter site change between Fe and Si) or to the strong hybridization
effects. It must be recalled here that the properties of the 2-3-5 silicides are highly 
anisotropic \cite{r15} and the over all shape of the temperature dependence of $\rho$(T) of 
a polycrystal can easily be influenced by the anisotropy. The important aspect is the
the low temperature  $\rho$ data on an expanded scale as shown by the inset (b). 
From this inset one can clearly see the non-monotonic 
dependence of $\rho$(T) with a log~T dependence from 9 to 25~K. The low temperature 
$\rho$ data below 30~K display a typical Kondo lattice response with a minimum at 25~K 
followed by a maximum  of $\rho$(T) at 6~K. Subsequently, $\rho$(T) falls
sharply below 4.5~K due to coherence.  A small kink in the resistivity data at 1.7~K
[shown in the panel(a)] indicates a possible magnetic transition at 1.7~K,
which is in agreement with the susceptibility data.
\subsection{Heat-capacity studies on Yb$_2$Fe$_3$Si$_5$}
\label{sec:HCND}
\noindent
The temperature dependent C$_p$  data from 0.4 to 25~K of
Yb$_2$Fe$_3$Si$_5$  is  shown in the top panel (a) of Fig.~3. The inset in this
panel describes the low temperature behaviour which emphasizes the sharpness of the
antiferromagnetic transition. The large
jump ($\sim$ 4~J/mol K) at the T$_N$ clearly shows bulk magnetic ordering
of Yb$^{3+}$ moments. The antiferromagnetic ordering temperature obtained from 
the heat capacity
measurements is in accordance with those obtained from the $\chi$ and
$\rho$ measurements. The specific heat data of Lu$_2$Fe$_3$Si$_5$ in the
normal state is also shown in the same panel. 

In the second panel of Fig.~3, the temperature dependence of C/T data are 
shown. Well below, T$_N$, these data can be nicely fitted with a linear
and a cubic term, C~=~$\gamma$~T~+~$\beta$T$^{3}$, as shown in the
inset. The cubic term in this case corresponds to the contribution of
antiferromagnetic magnons, since the phonons are completely negligible
in this temperature range. The linear term obtained from this extrapolation,
$\gamma$~=~0.4J/Yb-mol~K$^2$, can be considered as a lower limit for the
Sommerfeld coefficient, since the experimental data show an upward
curvature at the lowest limit of the measurement. This would indicate
freezing out of the magnons due to an anisotropy gap. In this case,
C/T~=~0.5~J/Yb-mol~K$^2$ at the lowest measured temperature gives the
upper limit for this Sommerfeld coefficient. The magnetic 
contribution to the heat capacity  of Yb$_2$Fe$_3$Si$_5$  (which is obtained 
after subtracting the C$_p$ data of Lu$_2$Fe$_3$Si$_5$) 
is shown in the third panel (c) of the Fig.~3. The estimated entropy from the experiment
is also shown in the same panel. The entropy is nearly temperature independent
from 7 to 12~K and reaches a value of 0.9~R~ln~2 (R is a gas constant) at 20~K, 
which establishes that the ground state is a well separated doublet. 
The heat capacity also shows a 
small negative curvature above 1.8~K to about 3~K. There are two
possibilities for the tail in C$_p$ above T$_N$: short range fluctuations or Kondo effect.
We would argue that the sharpness of the transition and the weak
T-dependence of C/T above T$_N$ are in disagreement with short range
correlations, whereas the Kondo scenario is supported by the resistivity
maximum at 6 K and the large $\gamma$ value. Moreover,
the entropy at T$_N$ [S$_{mag}$~=~S(T$_N$)]is found to be 0.39~R~ln~2, which strongly 
indicates that the ordered moment is heavily compensated by Kondo interactions. 
In the absence of short range order, the reduction in S$_{mag}$ at T$_N$
is due to the partial lifting of the two fold degeneracy of the ground
state above T$_N$ by Kondo effect. In such cases, one can show that
S$_{mag}$ = S$_K$(T$_N$/T$_K$), where S$_K$(t) is the entropy
of a single ion Kondo impurity at the normalized Kondo temperature,
t~=~T$_N$/T$_K$ \cite{r16}. The entropy has been calculated by 
Desgranges and Schotte using the Bethe ansatz for a spin-1/2 Kondo model
\cite{r17}. Sharp heat capacity anomaly at T$_N$ and the possible absence of short range
order facilitate the use of the theory \cite{r17} and with the experimental value of S$_{mag}$, 
we estimate a value of 5~K for the Kondo temperature of Yb$_2$Fe$_3$Si$_5$. However, accurate
estimation requires inelastic scattering data. The value of the entropy at 20~K is 
only 0.7~R which is much smaller than the theoretical value of 2.08~R [R~ln(2J+1) 
with J=7/2 for the free Yb$^{3+}$ ion]. This indicates that the CEF levels are
well splitted by the tetragonal symmetry.
\section{CONCLUSION}
\label{sec:CON}
To conclude, we have observed antiferromagnetic ordering and heavy electron
behaviour in Yb$_2$Fe$_3$Si$_5$ using resistivity, magnetization and
heat capacity studies. It is interesting to note that both Lu$_2$Fe$_3$Si$_5$ and 
Tm$_2$Fe$_3$Si$_5$ are superconducting at 6.2~K and 1.2~K, respectively, while
the Yb sample did not show superconductivity down to 100 mK. It should be
noted here that the superconducting Tm$_2$Fe$_3$Si$_5$ compound becomes 
reentrant antiferromagnet at 1.1~K and remains
in the same state down to 50 mK. We believe that the absence of superconductivity
in Yb sample is due to the Kondo effect, which is quite strong in suppressing
the superconductivity. The magnetic ordering temperature of Yb$_2$Fe$_3$Si$_5$ 
is also higher than that of the Tm compound which implies an ehanced exchange
in the former.  Here, it is the Fe-neighbors, which provide a very strong
d-f hybridization, and the Si-neighbors, which provide a significant
p-f hybridization. In view of the large effect of pressure on the properties
of Tm$_2$Fe$_3$Si$_5$, similar studies are essential on Yb$_2$Fe$_3$Si$_5$
to induce superconductivity via quenching of the Kondo effect. Such studies
are in progress and they will be reported elsewhere.

\begin{figure}
\caption{Temperature dependence of the inverse susceptibility (1/$\chi_{dc}$) of
Yb$_2$Fe$_3$Si$_5$ in a  field of 1~kOe from 1.9 to 300~K. 
The inset shows the low temperature ac $\chi$ data of
Yb$_2$Fe$_3$Si$_5$, which displays the antiferromagnetic transition at 1.7~K. 
The solid line is a fit to the Curie-Weiss relation (see text).
\label{fsus1}}
\end{figure}
\begin{figure}
\caption{Temperature dependence  of the resistivity ($\rho$) of
of Yb$_2$Fe$_3$Si$_5$ from 2 to 300~K. The inset (a) shows
the $\rho$ data from 1.4 to 5~K on an expanded scale whereas the
inset (b) shows $\rho$ data upto 50~K.  A small kink which
is marked by an arrow in the inset (a) indicates antiferromagnetic
ordering of Yb$^{3+}$ moments, which is in agreement with the $\chi$ data.
The solid line shown in the inset (b) is a fit to the Log T 
dependence (see text).
\label{fres1}}
\end{figure}
\begin{figure}
\caption{Plot of the heat-capacity (C$_p$) vs temperature (T) of
Yb$_2$Fe$_3$Si$_5$  from 0.3 to 25~K. The inset in the top panel (a) shows
the low temperature C$_p$ data on an expanded scale. Plot of C/T vs T
is displayed in the middle panel (b)which gives un upper limit for the value of
$\gamma$ as 0.5~J/Yb~mol~K$^2$ and the inset shows the extrapolated $\gamma$(0)
(0.4~J/Yb mol K$^2$) value using C/T vs T$^2$ plot. The bottom panel (c) 
depicts the magnetic contribution to the heat
capacity (after subtracting the lattice contribution estimated from
the C$_p$ data of Lu$_2$Fe$_3$Si$_5$) and the estimated entropy in this
temperature range.
\label{fcp1}}
\end{figure}

\begin{references}
\bibitem{r1} P. Rogl in {\it Handbook of Physics and Chemistry of Rare
Earths}, edited by K.A. Gschneidner, Jr. and L. Eyring (Elsevier Science
Publishers, North-Holland, Amsterdam, 1984), Vol. 7, pp 1-264.
\bibitem{r2} J. Leciejewicz and A. Szytula in {\it Handbook of Physics and
Chemistry of Rare Earths}, edited by K.A. Gschneidner, Jr. and L. Eyring
(Elsevier Science Publishers, North-Holland, Amsterdam, 1989), Vol. 12,
p. 133.
\bibitem{r3} H.F. Braun, J. Less Common Metals {\bf 100}, 105 (1984)
\bibitem{r4} R.N. Shelton in {\it Proc. Int. Conf. on Superconductivity
in d- and f- band Metals}, edited by W. Buckel and W. Weber
(Kernforshungszentrum, Karlsruhe, Germany 1982), p. 123.
\bibitem{r5} H.F. Braun, Phys. Letters {\bf 75A}, 386 (1980).
\bibitem{r6} H.F. Braun, C.U. Segre, F. Acker, M. Rosenberg, S. Dey,
U. Deppe, J. Magn. Magn. Mater. {\bf 25}, 117 (1981).
\bibitem{r7} A.R, Moodenbaugh, D.E. Cox, and H.F. Braun, Phys. Rev. B
{\bf 25}, 4702 (1981).
\bibitem{r8} C.B. Vining and R.N. Shelton, Phys. Rev. B {\bf 28},
2732 (1983).
\bibitem{r9} H.F. Braun in : {\it Ternary superconductors}, ed. by G.K.
Shenoy, B.D. Dunlap and F.Y. Fradin (North-Holland, Amsterdam, 1980),
p. 225.
\bibitem{r10} J.A. Gotaas, J.W. Lynn, R.N. Shelton, P. Klavins and
H.F. Braun, Phys. Rev. B {\bf 36}, 7277 (1987).
\bibitem{r11} S. Noguchi and K. Okuda, Physica B {\bf 194-196}, 1975 (1994).
\bibitem{r12} {\bf FULL PROF} X-ray powder diffraction program available at
Collaborative Computational Project Number 14 (CCP14) (www.ccp14.ac.uk).
\bibitem{r13} S. Ramakrishnan, S. Sundaram, R. S. Pandit and Girish
Chandra, J. of Phys. E {\bf 18}, 650 (1985).
\bibitem{r14} P.G. de Gennes, J. Phys. Radium {\bf 23}, 510 (1962).
\bibitem{r15} B. Becker, S. Ramakrishnan , A. A. Menovsky, G. J. Nieuwenhuys
and J. A. Mydosh, Phys. Rev. Lett. {\bf 78}, 1347 (1997). See also \cite{r11}.
\bibitem{r16} K. Satoh, T. Fujita, Y. Maeno, Y. Uwatoko and H. Fujii,
J. Phys. Soc. Jpn. {\bf 59}, 692 (1990)
\bibitem{r17} H. U. Desgranges and K. D. Schotte, 
Phys. Lett. {\bf 91A}, 240 (1982).
\end{references}
\end{document}